\documentclass[aps,pra,twocolumn,floatfix,superscriptaddress]{revtex4-1}
\usepackage{dcolumn,amsmath}
\usepackage{graphicx}
\usepackage{bm}
\usepackage{hyperref} 
\usepackage{natbib}

\begin{document}
\title{Saturated configuration interaction calculations for five-valent Ta and Db}
\date{\today}
\author{A. J. Geddes}
\author{D. A. Czapski}
\author{E. V. Kahl}
\author{J. C. Berengut}
\affiliation{School of Physics, University of New South Wales,
Sydney 2052, Australia}
\date{17 May 2018}
\begin{abstract}
Accurate atomic structure calculations of complicated atoms with 4 or more valence electrons begin to push the memory and time limits of supercomputers. This paper presents a robust method of decreasing the size of \textit{ab initio} configuration interaction and many-body perturbation theory calculations without undermining the accuracy of the resulting atomic spectra. Our method makes it possible to saturate the CI matrix in atoms with many valence electrons. We test our method on the five-valence-electron atom tantalum and verify the convergence of the calculated energies. We then apply the method to calculate spectra and isotope shifts of tantalum's superheavy analogue dubnium. Isotope-shift calculations can be used to predict the spectra of superheavy isotopes which may be produced in astrophysical phenomena.
\end{abstract}

\maketitle
\section{Introduction}
Configuration interaction and many-body perturbation theory (CI+MBPT) has been developed as an \textit{ab initio} method for accurately predicting the spectra of atoms with up to a few valence electrons. CI+MBPT was first introduced to calculate the spectra of neutral thallium~\cite{DzubaCIMBPT1996}. Since its development, it has been successful in accurately reproducing the spectra of atoms with 1-3 valence electrons (see e.g. \cite{Torretti2017,BerengutCI2006,BerengutMg2005}). 

Conversely, the CI+MBPT method has presented some limitations when applied to atoms with more than four valence electrons \cite{BerengutAlpha2011,berengut11pra1,Ongdwarf2013,SavukovCIMBPT2015}. The underlying method of CI+MBPT treats the valence-valence electron correlations using CI and incorporates core-valence interactions using MBPT. Recently the particle-hole CI+MBPT formalism has allowed important hole-excitations to be treated using CI \cite{BerengutHgHoles2016}, however there are computational constraints on its implementation. The size of the Hamiltonian matrix generated in the CI routine increases dramatically with the addition of more electrons which in turn places heavier demands on computational recourses. Moreover, accurate spectra calculations require the CI basis set to be large enough for saturation of the CI wavefunction. Consequently, the time and memory needed to diagonalise the CI matrix can exceed the supercomputer capacities typically available to atomic physicists before saturation is achieved.

This paper presents a way of minimising the computational resources needed for CI calculations. The ``emu CI'' method (Sec.~\ref{sec:emu}) can be implemented directly within existing CI+MBPT programs without sacrificing accuracy of results. We have tested the convergence of the method on the 5-valence-electron system tantalum with atomic number $Z=73$, and compared our results with usual CI+MBPT and existing experimental data. We then calculate the spectra and isotope shift constants for tantalum's superheavy analogue dubnium ($Z=105$) and estimated uncertainties. The method we present may allow for more accurate structure calculations in atoms with many valence electrons without the need for major modifications to existing code.

To date, it has been assumed that increasing the CI basis set will result in the calculated atomic spectra 
converging to a value close to the experimental data. Due to computational limits, the assumption that CI converges at large basis sets has not been tested. The emu CI modification allows for the CI basis set to be increased to very large sizes. 

Theoretical predictions of spectra and ionisation energies for super-heavy atoms will be important for experimental work in the future. The ionisation potential for lawrencium has been measured using the surface ionisation technique where a Lr$^+$ ion is formed on a high temperature surface and then selectively mass-separated from other by-products \cite{natureLr}. Theoretical predictions of the transition energies of Lr and Lr$^+$ were used to interpret the experimental results and extract the ionisation potential from the data. Alternatively, nobelium has been experimentally characterised using laser resonance spectroscopy \cite{natureNo}. Accurate predictions of No transition energies and strengths were used to locate transitions whilst avoiding broad wavelength scans, and also as a tool for comparison. The aforementioned methods are expected to be applied to even heavier elements of $Z \geq 104$ making it necessary to perform theoretical calculations for super-heavy elements.  

\textit{Ab initio} calculations of isotope shifts in superheavy atoms ($Z > 100$) can aid the search for the nuclear island of stability. Nuclear shell theory predicts that superheavy elements with a magic number of neutrons $N=184$ and an atomic number $Z \geq 104$  will be more stable than their lighter isotopes \cite{oganessian2004,hamilton2013}. It is not possible to produce neutron deficient superheavy atoms by colliding lighter atoms because the smaller, stable atoms have neutron to proton ratios smaller than are necessary to produce stable superheavy atoms. Alternatively, it is possible that neutron-rich superheavy atoms can be created by $r$-process neutron capture in astrophysical events such as neutron star mergers \cite{Goriely2011}. As a consequence, evidence of neutron-rich superheavy atoms may be present in complex astrophysical spectra. This appears to be feasible as atoms with atomic number up to $Z=99$ have been identified in the spectra of Przybylski's star \cite{gopka2008}. It has been proposed in \cite{DzubaWebb2017} that it is possible to predict the spectra of stable superheavy atoms using the spectra of the neutron-deficient isotope produced on Earth combined with accurate isotope shift calculations. Therefore, it may be possible to search for stable superheavy atoms in astrophysical spectra.

In addition, improving the \textit{ab initio} methods for characterising complex atoms will capacitate a wider search for potential atomic clock candidates and in turn assist the search for physics beyond the Standard Model \cite{BerengutMarchenkoAlpha}. Certain transitions in atomic clocks are highly sensitive to variations in the fine structure constant $\alpha$ \cite{BerengutHCI,BerengutAlpha2011}. In addition, interactions between dark matter and ordinary matter can present itself as variation in $\alpha$. Therefore, atoms sensitive to $\alpha$ variation may also be useful for putting limits on the existence of certain types of dark matter.

Nevertheless, many atomic clock candidates that are both favourable for atomic clocks and sensitive to new physics also have complicated valence configurations with many electrons. For example, the atomic clock candidate Ho$^{14+}$ \cite{DzubaHo14} has seven valence electrons, five of which reside in a $f$ shell. The spectra of Ho$^{14+}$ is therefore difficult to theoretically characterise due to the time and memory required to perform the calculation. Therefore the CI+MBPT method must be modified in order to overcome the challenges of atomic structure calculations in many-valence electron atoms. 

\section{Method}

Tantalum has a ground state configuration of [Xe]$4f^{14}5d^36s^2$. The CI+MBPT calculation commences with a Dirac-Hartree-Fock (DHF) treatment of the atom. The atom is modelled as a collection of single electrons in a nuclear potential with an additional mean potential $V^{N_{DF}}$ arising from all other surrounding electrons. The single electron wavefunctions $|m\rangle$ and energies $\epsilon_m$ are found by solving the DHF equations:
\begin{gather*}
\hat h^{DF} |m\rangle = \epsilon_m |m\rangle \\
\hat h^{DF} = c\, \boldsymbol{\alpha}\cdot\boldsymbol{p} + (\beta-1)m_e c^2 - \frac{Z}{r}+V^{N_{DF}}
\end{gather*}
The number of electrons included in the DHF calculation, $N_{DF}$, is typically chosen in the range from $N_e-M$~\cite{dzuba05pra1} up to $N_e$, where $N_e$ is the number of electrons and $M$ is the number of valence electrons. Ideally, the result of a fully converged CI+MBPT calculation will be the same regardless of which $V^{N_{DF}}$ is chosen. However, all CI basis sets are truncated, hence it is advantageous to choose the $V^{N_{DF}}$ that converges the most rapidly. An initial DHF calculation made it clear that the $5d^36s^2$ state is well separated from the energies of the lower filled $f$ orbitals, hence it is sensible to place the Fermi energy just above the filled $4f^{14}$ shell. The $V^N$, $V^{N-1}$ and $V^{N-5}$ potentials were tested using small CI calculations. The $V^{N-1}$ potential produced results that were more representative of the experimental spectra than the two other potentials and thus was used in calculations throughout this work. In a similar manner, the DHF valence configurations $5d^36s$, $5d^36p$ and $5d^26s6p$ were examined and consequently $5d^36s$ was selected. Note that at DHF level we treat open shells by simply scaling the potential due to the filled shell (i.e. the $5d^3$ potential is just $3/10$ that of a filled $5d^{10}$ shell).

Once the potential $V^{N_{DF}}$ is determined by solving the DHF equations, a single particle basis set $|i\rangle$ is constructed out of B-splines which have been diagonalised over $\hat h^{DF}$~\cite{johnson88pra,shabaev04prl}. This basis set contains core, valence and virtual states. States in the lower continuum are discarded.

\subsection{Configuration Interaction}
Slater determinants are constructed from the single-electron B-spline basis set $|i\rangle$. Configuration state functions (CSFs) $|I\rangle$ with a well defined total angular momentum $J$ and projection $M$ are formed by taking all Slater determinants with a given $M$ and diagonalising over the operator $\boldsymbol{\hat J}^2$. The many-electron wavefunction that will form the basis set for the CI calculation is formed from a linear combination of CSFs.

Next, the many-electron Hilbert space is split into two subspaces, both of which can be accounted for in different ways. The first subspace $Q$ contains all states with excitations from the [Xe]$4f^{14}$ core. The core-valence interactions are assumed to be small and hence we can treat $Q$ with MBPT as described in the next subsection. The complement of $Q$ is $P$. The states in subspace $P$ represent states with fully filled cores, hence the core can be frozen and the CSFs need only contain the valence electrons. This is valid if the valence and core electrons are well separated in energy, which in the case of Tantalum has been confirmed in the initial DHF calculation. All CSFs in $P$ are included directly in the CI method and describe valence-valence interactions. The effective CI Hamiltonian can be written using the Feshbach operator \cite{LindgrenMorrison}:
\begin{equation}
(PHP + \Sigma(E)) \Psi_P = E \Psi_P 
\label{eq:ham}
\end{equation}
where $PHP$ is the projection of the exact CI Hamiltonian onto the subspace $P$, $\Psi_P$ is the CI wavefunction, $E$ is the energy eigenvalue, and $\Sigma(E)$ is an operator that can be treated using MBPT. 

In principle, the subspace $P$ has an infinite number of dimensions as there are an infinite number of configurations into which the valence electrons can arrange themselves. Electrons are excited from an appropriate set of valence-electron leading configurations. In order to make the numerical CI treatment of $P$ computationally possible, it is necessary to constrain the number of valence electron configurations and hence CSFs which are included in CI. In our calculations, we limit the configurations to those created by one and two-electron excitations from the leading valence configurations $5d^36s^2$, $5d^46s$, $5d^5$, $5d^36s6p$ and $5d^26s^26p$. We have omitted additional leading configurations such as $5d^3 6p^2$ as they had little effect on the resulting energy levels but increased the time taken for the CI calculation significantly. Furthermore we truncate the single-particle basis set. In tantalum, the largest basis set we could use for a standard CI+MBPT calculation was $13spdf$. That is, we allow a valence basis with orbitals 6 -- 13$s$, 6 -- 13$p$, 5 -- 13$d$, and 5 -- 13$f$ (note that the higher energy waves are not spectroscopic since the B splines only extend spatially to $45$ Bohr radii). The number of CSFs corresponding to these configurations for a $J = 5/2$, odd-parity calculation is $N=244752$. The Hamiltonian is symmetric, so storing this matrix in memory at double precision requires 240~GB. We solve for the lowest eigenstates using the Davidson method~\cite{davidson75jcp} implemented by~\cite{stathopoulos94cpc}.

\subsection{Emu Configuration Interaction}
\label{sec:emu}

The next step in improving the CI method is to reduce the number of elements in the CI matrix without compromising the quality of our calculation. A CI matrix with fewer elements will require less memory and time to diagonalise.
Our strategy is similar that presented in \cite{DzubaSmallside2017}. The idea is that while higher-energy configurations contribute to the lower-energy levels of interest, this contribution is not strongly affected by interactions between the higher-energy configurations themselves. Ref.~\cite{DzubaSmallside2017} treated these higher-energy configurations using second-order perturbation theory in an implementation known as CIPT (CI and Perturbation Theory). Our approach is different in that we generate a full configuration interaction matrix but with matrix elements between these higher-energy CSFs set to zero.

In order to achieve this, we first need to ensure that all important CSFs that contribute strongly to the low-energy levels of interest are situated on one side of the CI matrix. There are $N_\textrm{small}$ of these important, typically lower-energy, CSFs. Any interactions between the important ($I \leq N_\textrm{small}$) and less important, typically higher-energy ($N_\textrm{small} \leq I \leq N$) CSFs will be included in the CI calculations as they will have a large effect on the lower-energy eigenstates. Conversely, the interaction between one higher state and a second higher state should have a negligible contribution to the overall CI wavefunction. All higher-higher interactions are therefore set to 0.

\begin{figure}[tb]
\centering
\includegraphics[width=7cm]{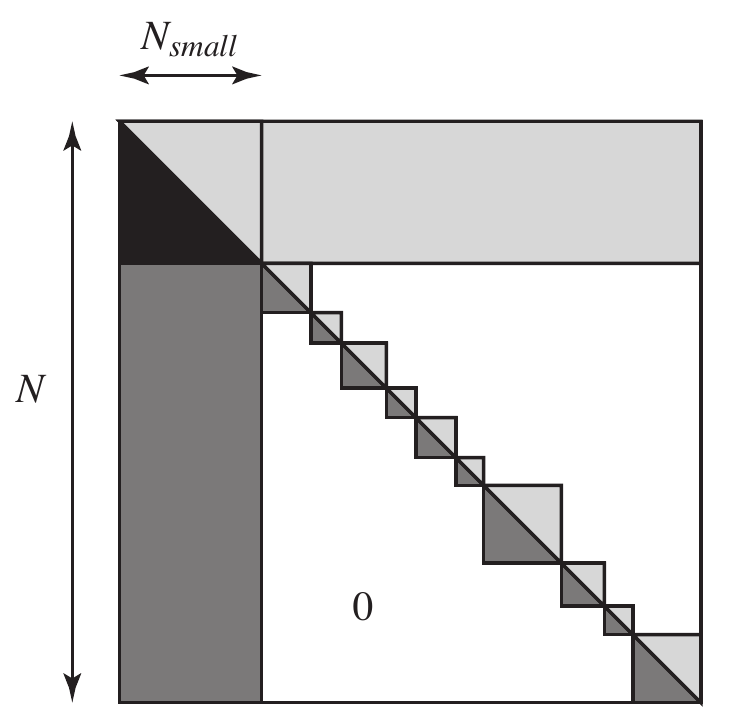}
\caption{Schematic depiction of the emu CI method (resembling an emu's footprint). The most important configuration state functions are in the upper left corner; the shaded areas become our effective CI matrix. In each square along the diagonal are matrix elements between CSFs belonging to the same relativistic configuration. The unshaded area represents interactions between high energy states which are set to 0.}
\label{fig:small}
\end{figure}

In the Figure \ref{fig:small} schematic, we see the lower-lying states are on the left of the matrix. Usual CI is represented by the black triangle (only the lower triangle is actually stored since the CI matrix is real and symmetric). The dark grey squares are also calculated and stored explicitly. The rectangle on the left hand side of Figure \ref{fig:small} represents the interactions between `lower' and `higher' CSFs. We also calculate elements in the squares along the diagonal. Each of these squares represents matrix elements between CSFs corresponding to the same relativistic configuration. The matrix elements between higher states are shown as the un-shaded areas and are set to zero. Therefore, the shaded area becomes our effective CI matrix.

The emu CI approach differs from the CIPT approach \cite{DzubaSmallside2017} in several respects. Firstly, CIPT treats the interaction of higher-energy configurations with the main configurations at second order in perturbation theory, while we treat them at all orders. Furthermore, in the emu CI approach we include the configurations as CSFs, which obey exact symmetries ($\boldsymbol{\hat J}^2$ and $\hat J_z$). Another important difference is that the CIPT method uses the configuration-average energy in the energy denominators of perturbation theory. In particular this means that the perturbation expansion is different for different targeted low-lying levels. In \cite{DzubaSmallside2017} this is dealt with by using an effective $E^{(0)}$ that averages the configuration energy over targeted levels. Emu CI avoids these issues because the higher-energy configurations are included directly in the CI calculation.

As we will see in Sec.~\ref{sec:Ta}, our emu CI calculation becomes saturated when our single particle basis includes orbitals up to $21spdf$. For the $J = 2.5$, odd-parity calculation $N = 952112$. However we find that we get good results by restricting the important configurations (on the small side of the rectangle) to those created by single excitations up to $21spdf$ and double excitations up to $6sp5d$, for which $N_\textrm{small} = 20462$. Thus the total number of matrix elements calculated and stored is reduced by a factor of $\sim 40$.

\subsection{Many Body Perturbation Theory}

Core-valence effects are considered using the MBPT method that was introduced in \cite{DzubaCIMBPT1996}. The subspace $Q$ includes all states where the core is not completely filled. The $\Sigma(E)$ operator introduced in Equation (\ref{eq:ham}) is dependent on energy, the projection of the $P$ states on $Q$ and vice-versa. It is possible to write $\Sigma(E)$ in terms of a perturbation expansion in the residual Coulomb interaction, which can be calculated using the diagrammatic technique described in \cite{LindgrenMorrison}. One, two and three body diagrams were taken into account up to the second order of perturbation theory using a large MBPT basis set of $30spdfgh$. Our implementation in the AMBiT code is detailed in \cite{BerengutCI2006}, with three-body diagrams introduced in \cite{BerengutTIII}.

\subsection{Isotope Shift}
When comparing the spectra of one isotope of a given atom with that of a second isotope, certain transitions will exhibit a small change in energy. The phenomena where atomic energy levels shift upon the addition of neutrons into the nucleus is known as isotope shift. The isotope shift between a pair of isotopes with mass numbers $A$ and $A'$ can be expressed as 
\begin{equation}
\delta \nu_{A,A'} = \nu_A - \nu_{A'} = K \left(\frac{1}{m_A} - \frac{1}{m_{A'}} \right) + F \delta \langle r^2 \rangle_{A,A'}.
\label{eq:IS}
\end{equation}
The first term is the result of the mass shift, where $K$ is the mass shift constant and $m_A$, $m_{A'}$ are the masses of the $A$ and $A'$ nuclei. The mass shift describes the effect of the motion of the nucleus with respect to the electrons, and how this changes when the mass of the nucleus increases. The second term of Equation \ref{eq:IS} corresponds to the field shift, where $F$ is the field shift constant and $ \delta \langle r^2 \rangle_{A,A'}$ is the change in the root-mean-squared (RMS) charge radius. The field shift incorporates the effect of the change in charge radius on the atomic energy levels, which is a consequence of the change in potential inside the nucleus.

Dubnium is a superheavy element with $Z=105$. Primarily, the isotope shift in heavy atoms such as dubnium is dominated by the field shift, as the nucleus is heavy enough that change in nuclear recoil becomes negligible. Therefore, it is only necessary to calculate $F$ from Equation \ref{eq:IS} when finding the isotope shift in dubnium. We determined the dubnium spectra by applying the approach used in our previous tantalum calculations. The DHF potential, CI leading configurations and basis set chosen where analogous to those used in tantalum, although increased by one principal quantum number to accommodate for dubnium's valence ground state of $6d^3 7s^2$. We can calculate the value of $F$ by modifying the charge radius in the CI+MBPT calculations. $F$ can be calculated from the change in frequency $d\omega$ with respect to RMS charge radius in the same way as presented in \cite{BerengutISOneVal,BerengutTIII}
\begin{equation}
F=\frac{d\omega}{\delta \langle r^2 \rangle}
\end{equation}

We calculated the dubnium energy spectra for five different nuclear radii spaced evenly either side of $R=1.5 A^{1/3}$ corresponding to $^{268}$Db. The transition energies were graphed against the matching $\langle r^2\rangle$ values to produce a linear relationship.  

\section{Results and Discussion}

\subsection{CI Convergence Testing in Ta}
\label{sec:Ta}

\begin{figure}[htb]
\centering
\includegraphics[width=9cm]{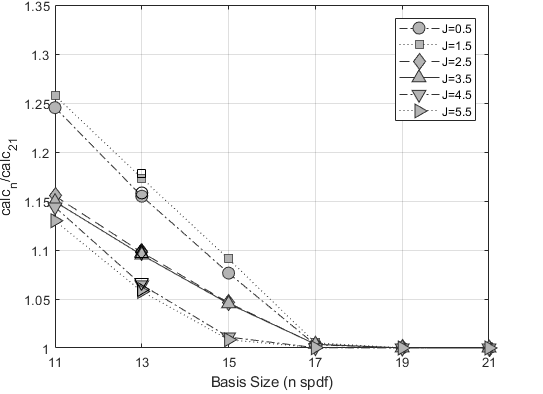}
\caption{Convergence of select even states when CI basis set is increased from $11spdf$ to $21spdf$. Open shapes denote results of the largest non-emu calculations performed.} 
\label{fig:evenCon}
\end{figure}

\begin{figure}[htb]
\centering
\includegraphics[width=9cm]{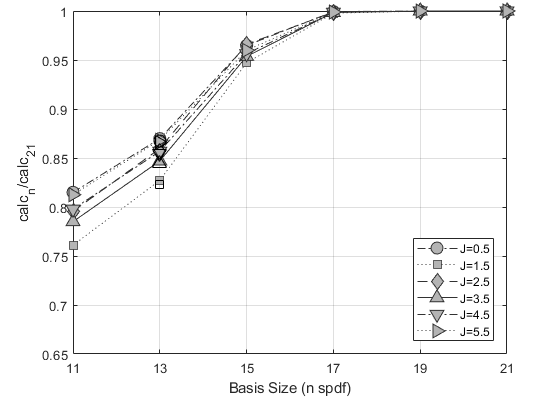}
\caption{Convergence of select odd states when CI basis set is increased from $11spdf$ to $21spdf$. Open shapes denote results of the largest non-emu calculations performed.} 
\label{fig:oddCon}
\end{figure}

The size of the CI basis set was increased from $11spdf$ in increments of two principal quantum numbers. The CI calculations performed with the full CI matrix were halted at $13spdf$ due to computational demands. It is apparent from Figures \ref{fig:evenCon} and \ref{fig:oddCon} that the full CI calculations did not meet convergence. 

\begin{figure}[htb]
\centering
\includegraphics[width=9cm]{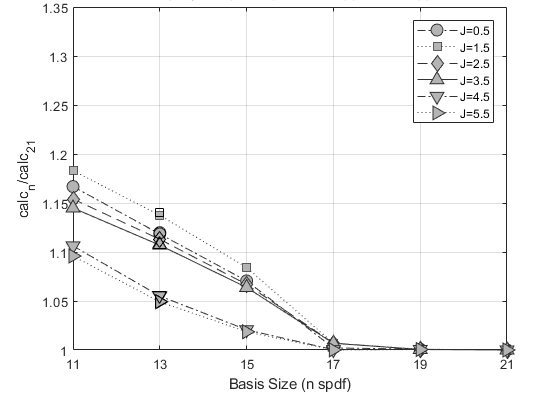}
\caption{Convergence of select even states when CI basis set is increased from $11spdf$ to $21spdf$ and MBPT is included. Open shapes denote results of the largest non-emu calculations performed.} 
\label{fig:evenConMBPT}
\end{figure}

\begin{figure}[htb]
\centering
\includegraphics[width=9cm]{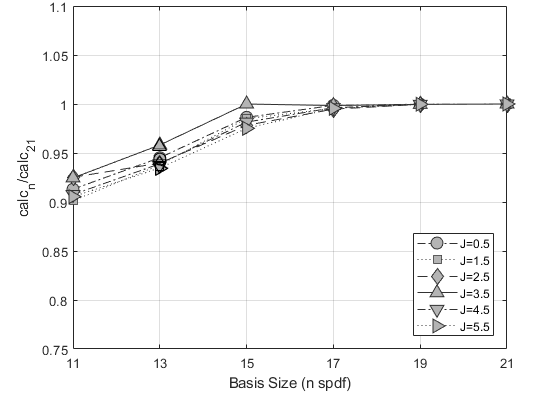}
\caption{Convergence of select odd states when CI basis set is increased from $11spdf$ to $21spdf$ and MBPT is included. Open shapes denote results of the largest non-emu calculations performed.} 
\label{fig:oddConMBPT}
\end{figure}

On the other hand, CI calculations with the emu CI method were performed for basis sets up to and including $21spdf$. The calculations were fully converged by $19spdf$ as illustrated in Figures \ref{fig:evenCon} and \ref{fig:oddCon}. It has been demonstrated in Figures \ref{fig:evenConMBPT} and \ref{fig:oddConMBPT} that the addition of MBPT did not alter the convergence of the odd or even states. 

The difference between the energies calculated with and without the emu CI method is small with respect to the change seen as the CI procedure converges. We compared the CI method truncated at $13spdf$ to an emu CI calculation with an equivalent basis set and found that the differences between the results were small, with the exception of a single state. The difference between the 13$spdf$ full CI and Emu CI was no more than 1\% of the non-emu calculation. Emu CI calculation was significantly less resource-intensive and provided comparable results, validating our assumption that interactions between higher energy states are negligible to a good approximation. Hence, for larger systems such as tantalum, it is far more advantageous to conduct calculations with much larger CI bases and a well-constructed emu CI approximation. 

Overall, the emu CI technique used in these calculations allowed convergence to be achieved both with and without MBPT and results did not differ significantly from calculations performed without the approximation, implying that all significant interactions contributing to the level positions have been accounted for. The largest CI+MBPT calculations conducted for tantalum were compared with literature values for the energy levels and have been presented in Tables \ref{table:TaEven} and \ref{table:TaOdd}. The six lowest energy solutions for each total angular momentum $J=\frac{1}{2},\ldots,\frac{11}{2}$ were calculated. We split our discussion into even and odd solutions.

Almost all of the lowest energy even parity levels found experimentally and presented in \cite{Arcimowicz2013,Windholz2018} were successfully identified in our calculations. The calculated levels were all within 14.5\% of the literature values, with some levels calculated within 1\% accuracy as depicted in Table \ref{table:TaEven}. Our calculations consistently overestimated the even states, with the average discrepancy being $\Delta E = -1583 \pm 1231$ cm$^{-1}$. 

The $E=43758$ cm$^{-1}$, $J=\frac{11}{2}$ literature level given was calculated using a semi-empirical parameter fit  presented in \cite{Arcimowicz2013} rather than being experimentally observed, however it is unclear whether this level exists.

Most of the lowest energy odd parity levels as given in \cite{Moore1958, Messnarz2003,Windholz2018} were identified in order and have been presented in table \ref{table:TaOdd}. All calculated odd levels were within $14\%$ of experimental values, with the majority of levels within $8\%$ of the experimental values, except the $E=15042$ cm$^{-1}$, $J=\frac{5}{2}$ level, which was within $16.5\%$. 
The best calculated odd level was also within 1\%. Our calculations consistently underestimated the odd states, with the average $\Delta E = 1773\pm714$ cm$^{-1}$. There is disagreement between the leading configurations calculated and those provided in literature (see \cite{Moore1958,vanDenBerg1952,Messnarz2003}), particularly with respect to levels with $5d^36s6p$ and $5d^26s^26p$ configurations. However, the tantalum spectrum is dense and there is a lot of mixing between states in both configurations.

In addition, the calculated values for the Tantalum energies (and also for dubnium) agree with those calculated by \cite{bryce} when we use a minimal emu CI approach (making $N_{small}$ very small), which is expected as the CIPT method used in \cite{bryce} is based off a very small CI calculation.

In both even and odd parities, some levels calculated appeared in a different energy order to those in the literature. We present the calculated energies in the energy order in which they were calculated. Calculations without MBPT were generally better for the even states, and with MBPT better for the odd states, however CI only was significantly worse for odd states.

\begin{table}[h!]
\label{table:even}
\centering
	\begin{ruledtabular}
    \begin{tabular}{ccccccc}
    ~ & ~ & \multicolumn{2}{c}{Calculated} & \multicolumn{2}{c}{Experimental} & ~ \\
    Config. & $J$ & $E$\tiny{(cm$^{-1}$)} & $g$ & $E$\tiny{(cm$^{-1}$)} & $g$ & $\Delta E$\tiny{(cm$^{-1}$)} \\ \hline
$	6s^2 5d^3	$	&	0.5	&	6215	&	2.4771	&	6049.3	&	2.454	&	-166	\\	
$	6s\,5d^4	$	&	0.5	&	10790	&	3.0605	&	9759	&	3.02	&	-1031	\\	
$	6s^2 5d^3	$	&	0.5	&	12564	&	1.0749	&	11792	&	1.116	&	-772	\\	
$	6s\,5d^4	$	&	0.5	&	22766	&	2.1834	&	20145	&	1.591*  &	-2621	\\	
$	6s\,5d^4	$	&	0.5	&	25441	&	0.45635	&	22236	&	1.056*  &	-3205	\\	
$	6s\,5d^4	$	&	0.5	&	29781	&	0.2274	&	26744	&	0.272*  &	-3037	\\	\hline
$	6s^2 5d^3	$	&	1.5	&	0		&	0.44895	&	0		&	0.447	&	0	\\	
$	6s^2 5d^3	$	&	1.5	&	6256	&	1.5488	&	6069	&	1.527	&	-187	\\	
$	6s^2 5d^3	$	&	1.5	&	10771	&	1.2928	&	9976	&	1.542	&	-795	\\	
$	6s\,5d^4	$	&	1.5	&	11708	&	1.6259	&	10950	&	1.407	&	-758	\\	
$	6s^2 5d^3	$	&	1.5	&	16846	&	1.2086	&	15904	&	1.199	&	-943	\\	
$	6s\,5d^4	$	&	1.5	&	24093	&	0.88567	&	21381	&	1.02	&	-2712	\\	\hline
$	6s^2 5d^3	$	&	2.5	&	2024	&	1.0353	&	2010	&	1.031	&	-14	\\	
$	6s^2 5d^3	$	&	2.5	&	9457	&	1.5806	&	9253	&	1.58	&	-204	\\	
$	6s\,5d^4	$	&	2.5	&	12341	&	1.6392	&	11244	&	1.641	&	-1097	\\	
$	6s^2 5d^3	$	&	2.5	&	13560	&	1.2159	&	12866	&	1.214	&	-694	\\	
$	6s^2 5d^3	$	&	2.5	&	18194	&	0.88261	&	17224	&	0.882	&	-970	\\	
$	6s\,5d^4	$	&	2.5	&	24135	&	0.85629	&	21623	&	0.894	&	-2512	\\	\hline
$	6s^2 5d^3	$	&	3.5	&	4100	&	1.221	&	3963.9	&	1.218	&	-136	\\	
$	6s^2 5d^3	$	&	3.5	&	10452	&	0.91595	&	9705.4	&	0.912	&	-746	\\	
$	6s\,5d^4	$	&	3.5	&	13419	&	1.5776	&	12235	&	1.578	&	-1184	\\	
$	6s^2 5d^3	$	&	3.5	&	18372	&	1.1342	&	17383	&	1.125	&	-989	\\	
$	6s\,5d^4	$	&	3.5	&	23203	&	0.82314	&	20647	&	0.818	&	-2556	\\	
$	6s\,5d^4	$	&	3.5	&	25310	&	0.9913	&	22761	&	1.008	&	-2549	\\	\hline
$	6s^2 5d^3	$	&	4.5	&	5935	&	1.2928	&	5621.1	&	1.272	&	-314	\\	
$	6s^2 5d^3	$	&	4.5	&	11524	&	1.0573	&	10690	&	1.063	&	-834	\\	
$	6s\,5d^4	$	&	4.5	&	14636	&	1.5365	&	13352	&	1.533	&	-1284	\\	
$	6s^2 5d^3	$	&	4.5	&	16443	&	1.0092	&	15391	&	1.014	&	-1052	\\	
$	6s\,5d^4 $	&	4.5	&	23749	&	1.0718	&	21153	&	1.089	&	-2596	\\	
$	6s\,5d^4	$	&	4.5	&	26536	&	1.1779	&	23913	&	1.185	&	-2623	\\	\hline
$	6s^2 5d^3	$	&	5.5	&	16292	&	1.0909	&	15114	&	1.021	&	-1178	\\	
$	6s\,5d^4	$	&	5.5	&	25068	&	1.1614	&	22429	&	1.159	&	-2639	\\	
$	6s\,5d^4	$	&	5.5	&	28747	&	1.2373	&	26023	&	1.26	&	-2724	\\	
$	6s\,5d^4	$	&	5.5	&	33523	&	0.96034	&	29499	&	0.975*  &	-4024	\\	
$	6s\,5d^4	$	&	5.5	&	37606	&	1.0611	&	33064	&	1.051*  &	-4542	\\	
$	6s\,7s\,5d^3	$	&	5.5	&	47087	&	1.441	&	43758$^{\#}$   	&	   1.248*   	&	-3329	\\	    \end{tabular}
    \end{ruledtabular}
    \caption{Energies and $g$ factors of even parity states in tantalum. Calculation performed with the emu CI method using a CI basis of $21spdf$ and MBPT basis of $30spdfgh$. $^{\#}$ this energy was calculated using a parameter fit rather than being observed experimentally. * these g-factors were calculated rather than being measured experimentally.} \label{table:TaEven}
\end{table}

\begin{table}[h!]
\label{table:odd}
\centering
	\begin{ruledtabular}
    \begin{tabular}{ccccccc}
    ~ & ~ & \multicolumn{2}{c}{Calculated} & \multicolumn{2}{c}{Experimental} & ~  \\
    Config. & $J$ & $E$\tiny{(cm$^{-1}$)} & $g$ & $E$ \tiny{(cm$^{-1}$)} & $g$ & $\Delta E$ \tiny{(cm$^{-1}$)} \\ \hline
   $	6s\,6p\, 5d^3	$	&	0.5	&	16269	&	0.42066	&	18504	&	0.172	&	2236	\\	
$	6s^2 6p\, 5d^2	$	&	0.5	&	17540	&	1.7001	&	20340	&	1.956	&	2801	\\	
$	6s\,6p\, 5d^3	$	&	0.5	&	21696	&	-0.17182	&	23355	&	-0.32	&	1659	\\	
$	6s\,6p\, 5d^3	$	&	0.5	&	22766	&	1.0153	&	24516	&	2.888	&	1751	\\	
$	6s\,6p\, 5d^3	$	&	0.5	&	23160	&	1.4873	&	25512	&	0.028	&	2353	\\	
$	6s\,6p\, 5d^3	$	&	0.5	&	24921	&	2.8377	&	26866	&	2.65	&	1945	\\	\hline
$	6s\,6p\, 5d^3	$	&	1.5	&	15770	&	0.3067	&	17384	&	   --      	&	1614	\\	
$	6s\,6p\, 5d^3	$	&	1.5	&	17646	&	0.81803	&	19658	&	1.018	&	2012	\\	
$	6s^2 6p\, 5d^2	$	&	1.5	&	18068	&	0.90576	&	20772	&	0.812	&	2705	\\	
$	6s^2 6p\, 5d^2	$	&	1.5	&	19551	&	0.59109	&	21855	&	0.666	&	2304	\\	
$	6s\,6p\, 5d^3	$	&	1.5	&	22437	&	1.4473	&	24243	&	1.126	&	1807	\\	
$	6s\,6p\, 5d^3	$	&	1.5	&	23103	&	1.251	&	24739	&	1.62	&	1636	\\	\hline
$	6s^2 6p\, 5d^2	$	&	2.5	&	15042	&	0.76253	&	17994	&	0.732	&	2952	\\	
$	6s\,6p\, 5d^3	$	&	2.5	&	17681	&	0.91497	&	19178	&	0.851	&	1498	\\	
$	6s^2 6p\, 5d^2	$	&	2.5	&	18695	&	1.0618	&	21168	&	1.12	&	2472	\\	
$	6s\,6p\, 5d^3	$	&	2.5	&	19767	&	1.1827	&	22047	&	1.179	&	2280	\\	
$	6s^2 6p\, 5d^2	$	&	2.5	&	20999	&	1.107	&	23363	&	1.078	&	2364	\\	
$	6s\,6p\, 5d^3	$	&	2.5	&	23588	&	1.202	&	25181	&	1.239	&	1593	\\	\hline
$	6s^2 6p\, 5d^2	$	&	3.5	&	18788	&	1.0695	&	20560	&	1.194	&	1772	\\	
$	6s\,6p\, 5d^3	$	&	3.5	&	20120	&	1.0976	&	22380	&	1.06	&	2261	\\	
$	6s^2 6p\, 5d^2	$	&	3.5	&	21880	&	1.2439	&	23927	&	1.326	&	2047	\\	
$	6s\,6p\, 5d^3	$	&	3.5	&	22844	&	1.3077	&	24982	&	1.235	&	2138	\\	
$	6s^2 6p\, 5d^2	$	&	3.5	&	24616	&	1.2285	&	26586	&	1.356	&	1970	\\	
$	6s\,6p\, 5d^3	$	&	3.5	&	25198	&	1.3297	&	26960	&	1.223	&	1762	\\	\hline
$	6s\,6p\, 5d^3	$	&	4.5	&	21491	&	1.2497	&	22682	&	1.231	&	1191	\\	
$	6s\,6p\, 5d^3	$	&	4.5	&	23141	&	1.2258	&	25186	&	   --   &	2045	\\	
$	6s^2 6p\, 5d^2	$	&	4.5	&	23843	&	1.2593	&	25926	&	1.292	&	2083	\\	
$	6s\,6p\, 5d^3	$	&	4.5	&	26410	&	1.3991	&	27734	&	1.39	&	1324	\\	
$	6s\,6p\, 5d^3	$	&	4.5	&	27446	&	1.2928	&	28767	&	1.337	&	1321	\\	
$	6s\,6p\, 5d^3	$	&	4.5	&	28747	&	1.3134	&	30021	&	1.186	&	1274	\\	\hline
$	6s\,6p\, 5d^3	$	&	5.5	&	24175	&	1.3304	&	25009	&	1.302	&	834	\\	
$	6s\,6p\, 5d^3	$	&	5.5	&	26483	&	1.3073	&	27783	&	1.351	&	1300	\\	
$	6s\,6p\, 5d^3	$	&	5.5	&	28584	&	1.3819	&	30361	&	1.334	&	1777	\\	
$	6s\,6p\, 5d^3	$	&	5.5	&	32867	&	1.1795	&	33070	&	1.349	&	203	\\	
$	6s\,6p\, 5d^3	$	&	5.5	&	34392	&	1.1167	&	34716	&	   --   &	1422	\\	
$	6s\,6p\, 5d^3	$	&	5.5	&	35590	&	1.2031	&	35814	&	1.2		&	-874	\\

    \end{tabular}
	\end{ruledtabular}
    \caption{
Energies %(in cm$^{-1}$)
and $g$ factors of odd parity states in tantalum. Calculation performed with the emu CI method using a CI basis of $21spdf$ and MBPT basis of $30spdfgh$. Literature configurations are from \cite{Moore1958}, except those labelled ``*'', which are from \cite{vanDenBerg1952} (see \cite{Messnarz2003} for more detail).}\label{table:TaOdd}
\end{table}

\subsection{Spectra and Field Shift in Db}

The spectra of dubnium was calculated using a CI basis set of $22spdf$, which is one principal quantum number higher then the maximum calculation performed for its analogue tantalum. The convergence graphs for tantalum imply that a dubnium calculation of this size has also converged. With the same reasoning, we can assume that the dubnium spectra has been calculated to within 10\% accuracy for most levels. The calculated spectra displayed in Tables \ref{Db-even} and \ref{Db-odd} correspond to the isotope with $A=268$ as discovered in laboratories. 

The Breit interaction was included in the HF procedure and resulted in the spectra shifting by $\sim 1$\%. This was expected as the dubnium system is highly relativistic. 

We have found the field shift constants for transitions between the ground state and each state shown in Tables \ref{Db-even} and \ref{Db-odd}, regardless of whether the transition is allowed or not. This way, it is possible to calculate the $F$ values of any transition; this can be done by subtracting the $F$ values of the excited and ground states of interest:

\begin{equation}
F^{a\rightarrow b} = F^{g \rightarrow b} - F^{g \rightarrow a}.
\end{equation}
Here $a$ and $b$ are the states of interest and $g$ is the ground state of dubnium.

\begin{table}[h!]
\centering
	\begin{ruledtabular}
    \begin{tabular}{ccccccc}
    ~ & ~ & \multicolumn{2}{c}{No Breit} & \multicolumn{2}{c}{Breit} & ~  \\
    Config.  & $J$  & $E$\tiny{(cm$^{-1}$)} 		& $g$  & $E$\tiny{(cm$^{-1}$)}		& $g$		& $F$ \tiny{(cm$^{-1}$/fm$^{2}$) }\\ \hline
$	7s^2 6d^3	$	&	0.5	&	8548	&	2.1857	&	8465	&	2.1913	&	-0.22	\\	
$	7s^2 6d^3	$	&	0.5	&	17462	&	1.2546	&	17309	&	1.2471	&	-0.14	\\	
$	7s\,6d^4	$	&	0.5	&	21226	&	3.0095	&	21063	&	3.0168	&	-6.32	\\	
$	7s^2 7p^2 6d	$	&	0.5	&	32624	&	0.59559	&	32752	&	1.3432	&	1.30	\\	
$	7s\,6d^4	$	&	0.5	&	33293	&	1.975	&	33264	&	1.2258	&	-4.85	\\	
$	7s^2 8s\, 6d^2	$	&	0.5	&	39588	&	2.0556	&	39578	&	2.0494	&	2.67	\\	\hline
$	7s^2 6d^3	$	&	1.5	&	0	&	0.57139	&	0	&	0.56905	&	0.00	\\	
$	7s^2 6d^3	$	&	1.5	&	8009	&	1.3471	&	7935	&	1.3507	&	0.24	\\	
$	7s^2 6d^3	$	&	1.5	&	16292	&	1.1184	&	16137	&	1.1192	&	0.63	\\	
$	7s\,6d^4	$	&	1.5	&	22776	&	1.2916	&	22738	&	1.5028	&	-0.64	\\	
$	7s\,6d^4	$	&	1.5	&	23544	&	1.4652	&	23539	&	1.2781	&	-2.61	\\	
$	7s^2 6d^3	$	&	1.5	&	26961	&	1.093	&	26851	&	1.07	&	1.85	\\	\hline
$	7s^2 6d^3	$	&	2.5	&	5321	&	1.0468	&	5235	&	1.0466	&	0.23	\\	
$	7s^2 6d^3	$	&	2.5	&	14485	&	1.527	&	14338	&	1.5288	&	0.56	\\	
$	7s^2 6d^3	$	&	2.5	&	19361	&	1.1928	&	19186	&	1.1936	&	0.48	\\	
$	7s^2 6d^3	$	&	2.5	&	22081	&	0.96885	&	21941	&	0.96571	&	0.72	\\	
$	7s\,6d^4	$	&	2.5	&	25511	&	1.6239	&	25265	&	1.6249	&	-7.24	\\	
$	7s^2 7p^2 6d	$	&	2.5	&	31053	&	1.1542	&	31131	&	1.1542	&	2.79	\\	\hline
$	7s^2 6d^3	$	&	3.5	&	8812	&	1.151	&	8699	&	1.1527	&	0.36	\\	
$	7s^2 6d^3	$	&	3.5	&	15429	&	1.0022	&	15273	&	1.0002	&	0.68	\\	
$	7s^2 6d^3	$	&	3.5	&	23685	&	1.119	&	23509	&	1.1193	&	0.57	\\	
$	7s\,6d^4	$	&	3.5	&	27567	&	1.5378	&	27302	&	1.5394	&	-7.24	\\	
$	7s\,6d^4	$	&	3.5	&	35098	&	0.87896	&	34889	&	0.87801	&	-7.66	\\	
$	7s^2 7p^2 6d	$	&	3.5	&	36156	&	1.0741	&	36412	&	1.0735	&	4.02	\\	\hline
$	7s^2 6d^3	$	&	4.5	&	10782	&	1.1672	&	10671	&	1.1696	&	0.40	\\	
$	7s^2 6d^3	$	&	4.5	&	17216	&	1.104	&	17041	&	1.1029	&	0.65	\\	
$	7s^2 6d^3	$	&	4.5	&	25646	&	1.0877	&	25382	&	1.0863	&	0.91	\\	
$	7s\,6d^4	$	&	4.5	&	29560	&	1.4834	&	29284	&	1.4856	&	-7.04	\\	
$	7s\,6d^4	$	&	4.5	&	36756	&	1.1175	&	36508	&	1.1162	&	-7.69	\\	
$	7s^2 6d^2 7d	$	&	4.5	&	42217	&	1.0217	&	42217	&	1.0327	&	2.57	\\	\hline
$	7s^2 6d^3	$	&	5.5	&	22826	&	1.0911	&	22637	&	1.0911	&	0.91	\\	
$	7s\,6d^4	$	&	5.5	&	39995	&	1.17	&	39703	&	1.1699	&	-7.77	\\	
$	7s\,6d^4	$	&	5.5	&	46566	&	1.1988	&	46214	&	1.2003	&	-7.54	\\	
$	7s^2 6d^2 7d	$	&	5.5	&	50506	&	1.0822	&	50310	&	1.0351	&	-0.33	\\	
$	7s\,6d^4	$	&	5.5	&	50937	&	1.0444	&	50728	&	1.0898	&	-4.01	\\	
$	7s^2 6d^2 8d	$	&	5.5	&	55446	&	1.2021	&	55407	&	1.1936	&	-0.05	\\

    \end{tabular}
    \end{ruledtabular}
    \caption{Energies and field shift constants of even parity states in Dubnium. Calculations were performed with the emu CI method with a CI basis of $22spdf$ and MBPT basis of $31spdfgh$. Energies and g-factors given are for $^{268}$Db.}\label{Db-even}
\end{table}

\begin{table}[!t]
\centering
	\begin{ruledtabular}
    \begin{tabular}{ccccccc}
    ~ & ~ & \multicolumn{2}{c}{No Breit} & \multicolumn{2}{c}{Breit} & ~ \\
    Final conf.	 & $J$  & $E$\tiny{(cm$^{-1}$)} 		& $g$  & $E$\tiny{(cm$^{-1}$)} 		& $g$		&  $F$\tiny{(cm$^{-1}$/fm$^{2}$) }  \\ \hline
    $	7s^2 7p\, 6d^2	$	&	0.5	&	10870	&	1.1843	&	11003	&	1.1942	&	1.49	\\	
$	7s^2 7p\, 6d^2	$	&	0.5	&	13551	&	1.138	&	13632	&	1.1272	&	1.62	\\	
$	7s^2 7p\, 6d^2	$	&	0.5	&	19921	&	0.11736	&	19992	&	0.11595	&	1.77	\\	
$	7s\,7p\, 6d^3	$	&	0.5	&	25971	&	1.3143	&	25957	&	1.3365	&	-1.73	\\	
$	7s\,7p\, 6d^3	$	&	0.5	&	28417	&	0.52521	&	28393	&	0.50336	&	-2.29	\\	
$	7s\,7p\, 6d^3	$	&	0.5	&	34029	&	2.5749	&	33986	&	2.6699	&	-3.81	\\	\hline
$	7s^2 7p\, 6d^2	$	&	1.5	&	9135	&	0.66478	&	9300	&	0.66361	&	2.28	\\	
$	7s^2 7p\, 6d^2	$	&	1.5	&	13620	&	1.1058	&	13708	&	1.1059	&	1.70	\\	
$	7s^2 7p\, 6d^2	$	&	1.5	&	17795	&	0.84716	&	17861	&	0.84603	&	0.01	\\	
$	7s^2 7p\, 6d^2	$	&	1.5	&	20082	&	1.2897	&	20137	&	1.2872	&	0.64	\\	
$	7s\,7p\, 6d^3	$	&	1.5	&	21621	&	0.37873	&	21656	&	0.38243	&	-3.06	\\	
$	7s^2 7p\, 6d^2	$	&	1.5	&	24999	&	1.257	&	24980	&	1.2568	&	1.47	\\	\hline
$	7s^2 7p\, 6d^2	$	&	2.5	&	4281	&	0.74799	&	4484	&	0.74789	&	3.04	\\	
$	7s^2 7p\, 6d^2	$	&	2.5	&	13231	&	1.0701	&	13323	&	1.0703	&	2.70	\\	
$	7s^2 7p\, 6d^2	$	&	2.5	&	16435	&	1.0884	&	16507	&	1.0888	&	1.32	\\	
$	7s^2 7p\, 6d^2	$	&	2.5	&	21069	&	1.1416	&	21108	&	1.1422	&	1.61	\\	
$	7s\,7p\, 6d^3	$	&	2.5	&	23729	&	0.92188	&	23725	&	0.92149	&	-4.29	\\	
$	7s^2 7p\, 6d^2	$	&	2.5	&	25872	&	1.0711	&	25851	&	1.0711	&	1.06	\\	\hline
$	7s^2 7p\, 6d^2	$	&	3.5	&	13571	&	1.0212	&	13663	&	1.0212	&	3.15	\\	
$	7s^2 7p\, 6d^2	$	&	3.5	&	19999	&	1.0986	&	20057	&	1.1004	&	2.52	\\	
$	7s^2 7p\, 6d^2	$	&	3.5	&	23096	&	1.2232	&	23114	&	1.2229	&	2.76	\\	
$	7s^2 7p\, 6d^2	$	&	3.5	&	25560	&	1.1551	&	25556	&	1.157	&	1.01	\\	
$	7s^2 7p\, 6d^2	$	&	3.5	&	26378	&	1.1649	&	26351	&	1.1638	&	0.04	\\	
$	7s\,7p\, 6d^3	$	&	3.5	&	28342	&	1.1395	&	28292	&	1.1391	&	-2.92	\\	\hline
$	7s^2 7p\, 6d^2	$	&	4.5	&	19425	&	1.1273	&	19477	&	1.1285	&	2.94	\\	
$	7s^2 7p\, 6d^2	$	&	4.5	&	25494	&	1.1381	&	25481	&	1.1383	&	2.30	\\	
$	7s^2 7p\, 6d^2	$	&	4.5	&	29055	&	1.1532	&	29014	&	1.153	&	1.71	\\	
$	7s\,7p\, 6d^3	$	&	4.5	&	31909	&	1.2446	&	31819	&	1.2456	&	-5.19	\\	
$	7s^2 7p\, 6d^2	$	&	4.5	&	35950	&	1.1412	&	35868	&	1.143	&	1.56	\\	
$	7s\,7p\, 6d^3	$	&	4.5	&	37844	&	1.2468	&	37727	&	1.2471	&	-1.79	\\	\hline
$	7s^2 7p\, 6d^2	$	&	5.5	&	31024	&	1.2383	&	30951	&	1.2392	&	1.48	\\	
$	7s\,7p\, 6d^3	$	&	5.5	&	36032	&	1.287	&	35921	&	1.2885	&	-4.62	\\	
$	7s^2 7p\, 6d^2	$	&	5.5	&	39226	&	1.1384	&	39114	&	1.138	&	0.86	\\	
$	7s\,7p\, 6d^3	$	&	5.5	&	42679	&	1.3199	&	42488	&	1.3214	&	-5.61	\\	
$	7s\,7p\, 6d^3	$	&	5.5	&	44525	&	1.1485	&	44386	&	1.1471	&	-5.96	\\	
$	7s^2 6d^2 6f	$	&	5.5	&	46883	&	0.99927	&	46930	&	0.99914	&	3.94	\\

    \end{tabular}
    \end{ruledtabular}
    \caption{Energies and field shift constants of odd parity states in dubnium. Calculations were performed with the emu CI method with a CI basis of $22spdf$ and MBPT basis of $31spdfgh$. Energies and g-factors given are for $^{268}$Db.}\label{Db-odd}
\end{table}

\section{Conclusion}
We have confirmed that reducing the size of the CI matrix in a CI+MBPT calculation significantly decreases the time and memory resources needed for large computations involving many valence electrons without having a significant impact on the accuracy of the results when compared to a standard CI+MBPT calculation. Consequently, the basis set used in these large calculations can be increased until saturation of the CI matrix is reached. We have demonstrated this through calculating the spectra of the 5 electron system tantalum with an accuracy within 10\% and testing the convergence of the results. Finally, we have applied this method to predict the spectra and isotope shifts in dubnium, neither of which have been experimentally measured. 

\section{Acknowledgements}
We thank A. Borschevsky for initiating this work and reviewing the manuscript.
A. J. Geddes and E. V. Kahl are grateful for the support of the Australian Government Research Training Program Scholarship.
D. A. Czapski would like to thank L. Windholz for valuable insight regarding experimental results.

\bibliographystyle{apsrev}
\bibliography{AMBiT.bib}

\end{document}